\documentclass[lettersize,journal]{IEEEtran}
\usepackage{amsmath,amsfonts}
\usepackage{algorithmic}
\usepackage{algorithm}
\pdfoutput=1
\usepackage{array}
\usepackage[caption=false,font=normalsize,labelfont=sf,textfont=sf]{subfig}
\usepackage{textcomp}
\usepackage{stfloats}
\usepackage{url}
\usepackage{verbatim}
\usepackage{graphicx}
\usepackage{cite}
\usepackage{hyperref}
\usepackage{multirow} 
\usepackage{booktabs} 
\usepackage{placeins} 
\usepackage{adjustbox} 
\usepackage{tabularx} 
\usepackage{siunitx} 
\usepackage{amsmath}  
\usepackage[justification=centering]{caption}

\begin{document}

\title{Exploring Speaker Diarization with  Mixture of Experts}

\author{Gaobin Yang, Maokui He, Shutong Niu, Ruoyu Wang, Hang Chen, Jun Du,~\IEEEmembership{Senior Member,~IEEE,}
\thanks{This paper was produced by the IEEE Publication Technology Group. They are in Piscataway, NJ.}
\thanks{Manuscript received April 19, 2021; revised August 16, 2021.}}

\markboth{Journal of \LaTeX\ Class Files,~Vol.~14, No.~8, August~2021}%
{Shell \MakeLowercase{\textit{et al.}}: A Sample Article Using IEEEtran.cls for IEEE Journals}


\maketitle

\begin{abstract}
In this paper, we propose a novel neural speaker diarization system using memory-aware multi-speaker embedding with sequence-to-sequence architecture (NSD-MS2S), which integrates a memory-aware multi-speaker embedding module with a sequence-to-sequence architecture. The system leverages a memory module to enhance speaker embeddings and employs a Seq2Seq framework to efficiently map acoustic features to speaker labels. Additionally, we explore the application of mixture of experts in spkeaker diarization, and introduce a Shared and Soft Mixture of Experts (SS-MoE) module, to further mitigate model bias and enhance performance. 
Incorporating SS-MoE leads to the extended model NSD-MS2S-SSMoE. Experiments on multiple complex acoustic datasets, including CHiME-6, DiPCo, Mixer 6 and DIHARD-III evaluation sets, demonstrate meaningful improvements in robustness and generalization. The proposed methods achieve state-of-the-art results, showcasing their effectiveness in challenging real-world scenarios. 
\end{abstract}


\section{Introduction}
\IEEEPARstart{S}{peaker} diarization, which aims to determine the temporal boundaries of individual speakers within an audio stream and assign appropriate speaker identities, addresses the fundamental question of “who spoke when” \cite{review1}. It serves as a foundational component in numerous downstream speech-related tasks, including automatic meeting summarization, conversational analysis, and dialogue transcription \cite{review2}. Nevertheless, achieving robust diarization performance in practical settings remains a persistent challenge, primarily due to factors such as an unknown and variable number of speakers, acoustically adverse environments, and a high prevalence of overlapping speech segments.

Traditional clustering-based speaker diarization approaches \cite{CD1} typically consist of several sequential modules, including voice activity detection (VAD), speech segmentation, speaker representation extraction—such as i-vector \cite{ivector}, d-vector \cite{dvector}, and x-vector \cite{xvector}—speaker clustering \cite{AHC, COSINE, SC}, and subsequent re-segmentation procedures \cite{VBResegmentation1}. While such modular pipelines demonstrate considerable robustness across a variety of domains, they inherently struggle with overlapping speech segments, as each segment is constrained to a single speaker label due to the limitations of the clustering mechanism.

To overcome these limitations, neural-based diarization methods have been actively explored in recent years. Among them, End-to-End Neural Diarization (EEND) \cite{EEND1} represents a paradigm shift by integrating multiple diarization components—including voice activity detection, speaker embedding extraction, and speaker attribution—into a single, jointly optimized model. EEND reformulates speaker diarization as a multi-label frame-wise classification task, directly predicting speaker activities from audio features without relying on intermediate clustering. Building upon the original EEND framework, numerous improvements have been proposed to enhance its performance and applicability.Self-Attentive EEND (SA-EEND)\cite{EEND2}, which leverages the global modeling capability of self-attention mechanisms to improve the performance upper bound of EEND. To address the variable-number-of-speakers challenge, Encoder-Decoder-based Attractor Calculation EEND (EDA-EEND) \cite{EEND3} has been proposed. This method employs an additional attractor module to detect new speakers and integrates the attractor's outputs into the main network to guide the final diarization results. For extending EEND to online decoding scenarios, Xue et al. \cite{xue2021online} proposed an EEND variant with a speaker-tracking buffer, which aligns speaker labels across adjacent processing chunks using a tracking buffer. When processing long-duration audio, EEND faces significant computational and memory burdens due to the quadratic time complexity of attention mechanisms. To mitigate this issue, EEND-vector clustering (EEND-VC) \cite{wang2022incorporating} processes long audio in segmented chunks. Each chunk is independently decoded, and speaker-specific features from the same speaker are averaged along the time dimension and projected into a speaker embedding space. Finally, clustering algorithms are applied to the speaker embeddings to resolve speaker alignment across different chunks.

In parallel, Target-Speaker Voice Activity Detection (TS-VAD)
\cite{tsvad} has been proposed as a neural-based post-processing method to refine the outputs of traditional diarization systems. TS-VAD leverages prior speaker information to perform target-speaker detection and jointly models the activities of multiple speakers. Despite its widespread success in various applications \cite{Dihard-III, watanabe2020chime, cornell2023chime}, original TS-VAD still exhibits some limitations that have motivated numerous research efforts. A transformer-based TS-VAD \cite{wang2023target} architecture that handles variable numbers of speakers through representations with dynamic time and speaker dimensions. Meanwhile, an end-to-end target-speaker voice activity detection (E2E-TS-VAD) method \cite{wang2022incorporating} was proposed to jointly learn speaker representation and diarization refinement, achieving better performance than the original TS-VAD with clustering-based initialization. Seq2Seq-TSVAD \cite{seq2seq} adopted a sequence-to-sequence framework, demonstrating improved efficiency while maintaining accuracy. NSD-MA-MSE\cite{MAMSE} tackled the speaker embedding reliability issue through a memory-augmented neural network that dynamically refines speaker representations, thereby mitigating the domain gap between embedding extraction and neural network. 

To promote advances in speaker diarization under complex acoustic conditions, several international challenges have been organized to systematically benchmark algorithmic progress. Among them, the CHiME-7 Challenge and the DIHARD III Challenge are particularly notable. The CHiME-7 Challenge\cite{cornell2023chime},  introduced a main track focused on multi-speaker automatic speech recognition (ASR) centered on distant microphone conversational speech recorded under real-world conditions, where speaker diarization served as a critical front-end module to segment and organize speaker turns before transcription.  It utilized three datasets: CHiME-6\cite{watanabe2020chime}, DiPCo\cite{van2019dipco}, and Mixer 6\cite{brandschain2010mixer}. These datasets cover a wide range of challenging conversational scenarios, including multi-speaker dinner parties across kitchens, dining rooms, and living rooms, as well as interview sessions, telephone-style dialogues, and spontaneous dictations in controlled environments. Recordings are conducted with far-field microphone arrays and allow for natural speaker behaviors such as free movement, overlapping speech, and dynamic interaction patterns. Consequently, these datasets present significant challenges, including strong reverberation, background noise, severe speech overlap, and varying speaker counts. In a similar vein, the DIHARD-III Challenge\cite{Dihard-III} targets speaker diarization in highly diverse and challenging domains. Spanning 11 diverse domains, including clinical interviews, sociolinguistic fieldwork recordings, telephone conversations, YouTube videos, courtroom trials and so on. Both CHiME-7 and DIHARD-III have substantially contributed to pushing the limits of diarization technology, encouraging the development of systems that are more robust, generalizable, and capable of handling complex real-world scenarios. Despite substantial research efforts, many existing diarization systems still face challenges in achieving robust and generalized performance on these benchmarks\cite{cornell2023chime,}

The original TS-VAD framework, while effective, exhibits several notable limitations. First, its reliance on a BLSTM-based architecture results in high computational complexity, leading to slower inference speeds and substantial GPU memory consumption, particularly as the input sequence length increases. Second, TS-VAD typically employs a pre-trained extractor to generate speaker embeddings (such as i-vectors), but in real-world applications, these embeddings often suffer from degradation due to the absence of oracle speaker segments, thus compromising system robustness. Third, when deployed across diverse acoustic domains, TS-VAD models are susceptible to domain-specific biases, limiting their generalization capability and affecting performance consistency under mismatched conditions.To address these challenges, various improved methods have been proposed\cite{seq2seq, MAMSE, wang2023target, wang2022incorporating}. Nevertheless, existing solutions often mitigate only part of the issues, and a unified approach that simultaneously enhances efficiency, robustness, and generalization remains underexplored.

To address the aforementioned challenges, we propose a novel neural speaker diarization system using memory-aware multi-speaker embedding with sequence-to-sequence architecture (NSD-MS2S). Additionally, we explore the application of mixture of experts in spkeaker diarization, 
 and extend NSD-MS2S to NSD-MS2S-SSMoE. Consequently, the principal contributions of our study can be summarized as follows:
\begin{enumerate}[]
    \item NSD-MS2S seamlessly integrates the advantages of the Memory-Aware Multi-Speaker Embedding (MA-MSE) module and the Sequence-to-Sequence (Seq2Seq) architecture, achieving an efficient and powerful framework for speaker diarization. Then, we develop a simple yet effective feature fusion strategy, which significantly reduces the computational burden in the transformer's decoder without sacrificing diarization accuracy. To enhance the retrieval of multi-speaker embeddings from the memory module, we introduce a Deep Interactive Module (DIM) within the MA-MSE framework. By performing multi-scale feature fusion between acoustic features and speaker embedding basis vectors, DIM produces cleaner and more discriminative multi-speaker representations.
    
    \item To address the issue of model bias across different acoustic conditions, we further introduce a novel Shared and Soft Mixture of Experts (SS-MoE) module into the Seq2Seq-based diarization framework, resulting in the development of an enhanced system referred to as NSD-MS2S-SSMoE.

    \item We introduce a simple and effective parameter transfer strategy, where the pre-trained parameters from NSD-MS2S are migrated to initialize the NSD-MS2S-SSMoE model. This method accelerates the convergence of the SS-MoE enhanced system during training and reduces the overall training cost.

    \item Our proposed NSD-MS2S system achieved the first place in the main track of the CHiME-7 challenge. Furthermore, NSD-MS2S-SSMoE further improves single-model performance, achieving results comparable to the system fusion of NSD-MS2S on the CHiME-7 evaluation set, and attaining state-of-the-art performance on the DIHARD-III evaluation set.
\end{enumerate}

\section{Related work}
\subsection{Neural Speaker Diarization Using Memory-Aware Multi-speaker Embedding}

Traditional speaker diarization systems predominantly rely on clustering-based paradigms\cite{AHC, SC, LANDINI2022101254}. While effective in many scenarios, these methods exhibit significant challenges when encountering overlapping speech, due to the property of the clustering algorithm. To overcome these limitations, end-to-end neural diarization (EEND) approaches have been proposed, reframing diarization as a multi-label classification task. Representative methods such as EEND\cite{EEND1, EEND2, EEND3} directly predict frame-level speaker activity for all speakers simultaneously. Similarly, target-speaker voice activity detection (TS-VAD)\cite{tsvad, wang2023target, MAMSE, seq2seq,wang2023target} enhances speaker tracing by leveraging pre-acquired speaker embeddings to estimate speaker presence probabilities.
Neural Speaker Diarization Using Memory-Aware Multi-Speaker Embedding (NSD-MA-MSE) \cite{MAMSE} represents one of the state-of-the-art approaches among TS-VAD-based methods, which introduces a dedicated memory module designed to generate a set of speaker embeddings specifically for TS-VAD.

NSD-MA-MSE accepts a sequence of acoustic frames as input, represented by the matrix $\mathbf{X} = [\mathbf{x}_1, \mathbf{x}_2, \ldots, \mathbf{x}_T]$, where each $\mathbf{x}_t \in \mathbb{R}^{D'}$ corresponds to a $D'$-dimensional log-Mel filter-bank (FBANK) feature vector at time step $t$, and $T$ denotes the total number of frames in an utterance. This input is processed through four convolutional layers, which transform the raw acoustic features into higher-level representations. The resulting deep features are denoted as $\mathbf{F} = [\mathbf{f}_1, \mathbf{f}_2, \ldots, \mathbf{f}_T]$, with each $\mathbf{f}_t \in \mathbb{R}^{D}$ capturing a $D$-dimensional vector at frame $t$.
These frame-wise features are simultaneously used by the primary model and a memory-based module. To model speaker-specific characteristics, each deep feature is concatenated with a replicated set of speaker embeddings $\mathbf{E} = [\mathbf{e}_1, \mathbf{e}_2, \ldots, \mathbf{e}_N]$, where each $\mathbf{e}_n \in \mathbb{R}^L$ is an embedding for the $n$-th speaker. These embeddings are broadcast across all frames and are generated by the MA-MSE module. The combined representations (acoustic features and speaker embeddings) are then forwarded to a speaker detection component, which consists of a two-layer bidirectional LSTM with projection (BLSTMP) \cite{BLSTMP} to capture the temporal dependencies in speaker activity patterns.
Subsequently, the speaker-wise outputs from the SD module are aggregated and passed through a final one-layer BLSTMP to compute binary classification outputs for speech activity. The system generates $\mathbf{\hat{Y}} = (\hat{y}_{nt}) \in \mathbb{R}^{N \times T}$, where each $\hat{y}_{nt} \in [0, 1]$ denotes the probability that speaker $n$ is speaking at frame $t$.
\begin{figure}[t]
    \begin{center}
    \includegraphics[width=1.0\linewidth]{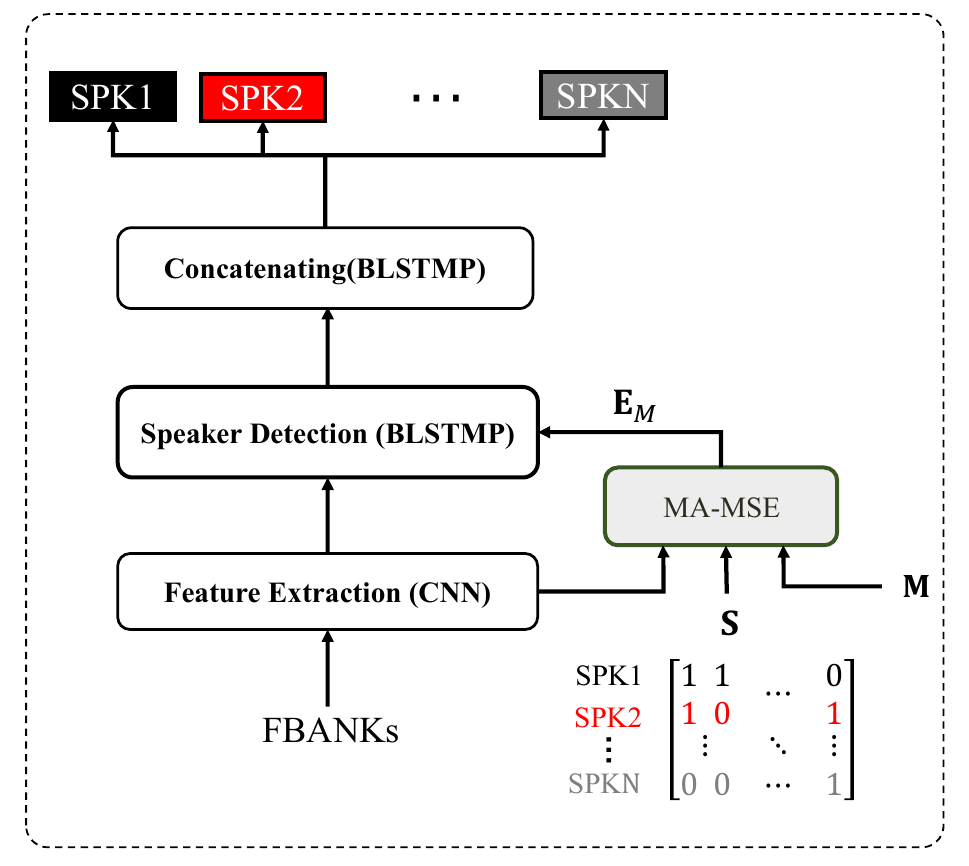}
    \caption{The architecture of neural speaker diarization network using memory-aware multi-speaker embedding.}
    \label{fig:mamse}
    \end{center}
\end{figure}

\subsection{Soft Mixture of Experts}
\begin{figure}[t]
    \begin{center}
    \includegraphics[width=1.0\linewidth]{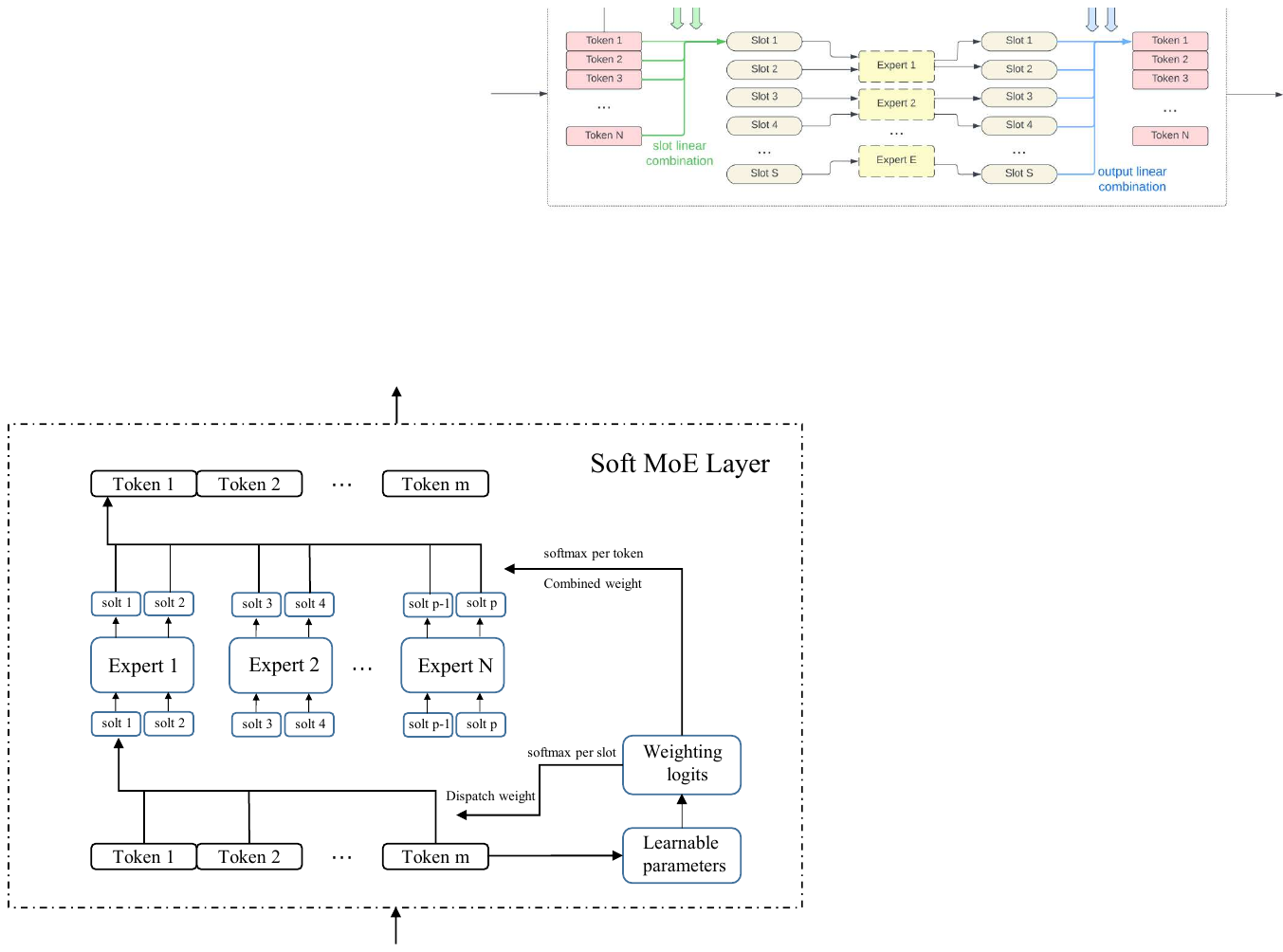}
    \caption{Soft MoE routing details.}
    \label{fig:softmoe}
    \end{center}
\end{figure}

The Soft MoE routing algorithm~\cite{puigcerver2023sparse} presents a token-to-slot assignment strategy that enables fully differentiable expert selection. Given input tokens $\mathbf{X} \in \mathbb{R}^{m \times d}$, where $m$ denotes the number of tokens and $d$ the token dimension, each expert in the mixture operates on $p$ virtual slots, parameterized by $\boldsymbol{\Phi} \in \mathbb{R}^{d \times (n \cdot p)}$. Here, $n$ is the number of experts, and the total number of slots is $n \cdot p$.

To compute the slot representations, a dispatch matrix $\mathbf{D} \in \mathbb{R}^{m \times (n \cdot p)}$ is first computed via softmax normalization over the columns of the similarity matrix $\mathbf{X}\boldsymbol{\Phi}$. Each slot representation is then obtained as a convex combination of all input tokens:
\begin{equation}
    \mathbf{D}_{ij} = \frac{\exp((\mathbf{X}\boldsymbol{\Phi})_{ij})}{\sum_{i'=1}^{m} \exp((\mathbf{X}\boldsymbol{\Phi})_{i'j})}, \quad 
    \tilde{\mathbf{X}} = \mathbf{D}^\top \mathbf{X}
\end{equation}

Each row of $\tilde{\mathbf{X}}$ is routed to a designated expert based on its slot index. The expert function $f_{\lfloor i/p \rfloor}$ then processes each slot independently to produce the intermediate output slots $\tilde{\mathbf{Y}}_i$.

A second softmax, applied row-wise to the same token-slot interaction scores, yields the combine matrix $\mathbf{C} \in \mathbb{R}^{m \times (n \cdot p)}$, which is used to reconstruct the final output tokens:
\begin{equation}
    \mathbf{C}_{ij} = \frac{\exp((\mathbf{X}\boldsymbol{\Phi})_{ij})}{\sum_{j'=1}^{n \cdot p} \exp((\mathbf{X}\boldsymbol{\Phi})_{ij'})}, \quad
    \mathbf{Y} = \mathbf{C} \tilde{\mathbf{Y}}
\end{equation}

This fully differentiable token-to-slot-to-token mechanism enables end-to-end training without hard routing decisions. In practical applications, a portion of the Transformer's feed-forward layers—typically the latter half—can be replaced by Soft MoE modules. The number of slots, rather than the number of experts, primarily influences the computational cost, making it a tunable parameter for balancing efficiency and performance. Typically, part of the Transformer's feed-forward layer can be replaced by the Soft-MoE block.

\section{Neural Speaker Diarization System Using Memory-Aware Multi-Speaker Embedding With Sequence-to-Sequence Architecture}
\subsection{Overview of Network}
\begin{figure*}[t]
    \begin{center}
    \includegraphics[width=1.0\linewidth]{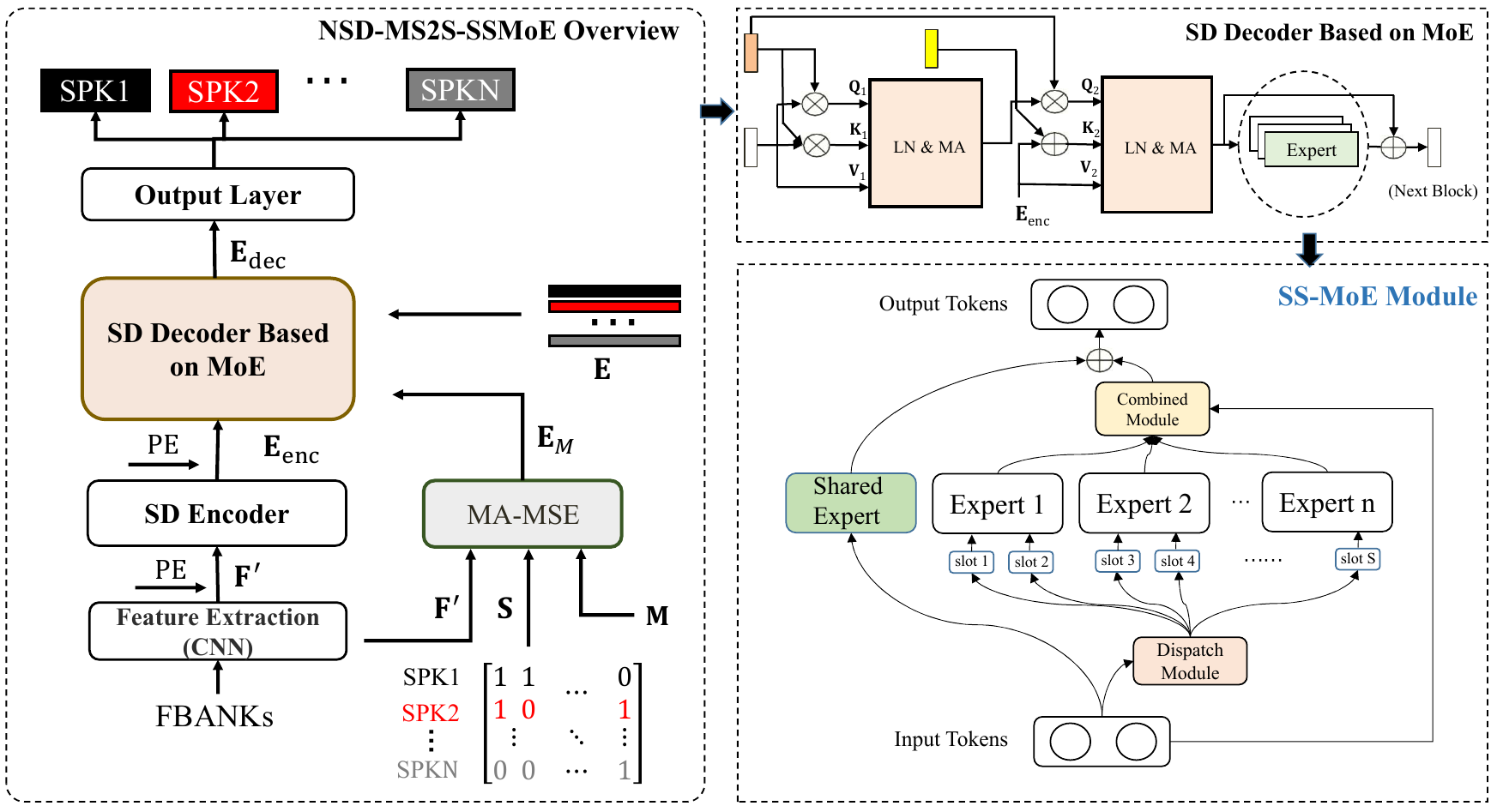}
    \caption{The proposed NSD-MS2S-SSMoE framework.}
    \label{fig:softmoe}
    \end{center}
\end{figure*}

The main network receives a sequence of acoustic features, denoted as $\mathbf{X} \in \mathbb{R}^{T \times F}$, where $T$ and $F$ represent the time steps and the dimensionality of log-Mel filter-bank features (FBANKs), respectively. These features are processed by convolutional layers to extract a set of deep features $\mathbf{F} \in \mathbb{R}^{C \times T \times \frac{F}{2}}$, which are then downsampled to produce $\mathbf{F'} \in \mathbb{R}^{T \times D}$, where $C$ and $D$ are the channel and feature dimensions, respectively.

The feature sequence $\mathbf{F'}$ is augmented with positional embeddings (PE) and passed through the speaker detection (SD) encoder, which consists of a stack of conformer blocks, yielding the encoded features $\mathbf{E}_{\text{enc}} \in \mathbb{R}^{T \times D}$.

Additionally, $\mathbf{F'}$ and the speaker mask matrix $\mathbf{S} \in \mathbb{R}^{N \times T}$ are input to the MA-MSE module, producing the MA-MSE embedding $\mathbf{E}_M \in \mathbb{R}^{N \times D_M}$, where $N$ is the number of speakers and $D_M$ is the dimensionality of the MA-MSE embedding. This embedding is concatenated with the i-vector to form the aggregated embedding $\mathbf{E}_A \in \mathbb{R}^{N \times D}$, which is further described in the latter.

The aggregate embedding $\mathbf{E}_A$, along with the decoder embedding $\mathbf{E}_D \in \mathbb{R}^{N \times D}$ and the encoded features $\mathbf{E}_{\text{enc}}$, are passed through the SD decoder, augmented with sinusoidal positional embeddings. This results in the decoded features $\mathbf{E}_{\text{dec}} \in \mathbb{R}^{N \times D}$, which are discussed in the latter. Finally, the output layer converts $\mathbf{E}_{\text{dec}}$ into posterior probabilities $\mathbf{\hat{Y}} = \left [ \hat{y}_\text{nt} \right ]_{N \times T}$, representing the voice activity probabilities for $N$ speakers.

\subsection{Speaker Detection Decoder}
\label{SD_Decoder}

The design of the speaker detection (SD) decoder is primarily inspired by \cite{seq2seq, dert}. It consists of multiple SD blocks that predict the voice activities of target speakers by considering cross-speaker correlations.

In the forward pass of an SD block, the decoder embedding $\mathbf{E}_D$ and the aggregate embedding $\mathbf{E}_A$ are processed through their respective multi-layer perceptrons (MLPs) to generate the within-block representations $\mathbf{E}_D^{Q_1}$, $\mathbf{E}_D^{K_1}$, $\mathbf{E}_D^{V_1}$, $\mathbf{E}_A^{Q_1}$, and $\mathbf{E}_A^{K_1}$, where $Q$, $K$, and $V$ represent the query, key, and value in the attention mechanism, respectively. All MLP layers, unless otherwise noted, map the input feature dimensions to $D$. The MLP structure is omitted for simplicity in Fig.\ref{fig:NSD-MS2S}(c).

To ensure that the decoder embedding includes speaker information while minimizing subsequent time and space overhead, the input features are fused without increasing the feature dimensions. This fusion can be expressed by the following equations:
\begin{equation}
    \mathbf{Q}_1 = \mathbf{\beta}_1 \times \mathbf{E}_D^{Q_1} + (1 - \mathbf{\beta}_1) \times \mathbf{E}_A^{Q_1} \nonumber
\end{equation}
\begin{equation}
    \mathbf{K}_1 = \mathbf{\beta}_2 \times \mathbf{E}_D^{K_1} + (1 - \mathbf{\beta}_2) \times \mathbf{E}_A^{K_1}
\end{equation}
\begin{equation}
    \mathbf{V}_1 = \mathbf{E}_D^{V_1} \nonumber
\end{equation}
where $\mathbf{\beta}_1$ and $\mathbf{\beta}_2$ are learnable parameters that allow the model to determine the most relevant information. The queries $\mathbf{Q}_1$, keys $\mathbf{K}_1$, and values $\mathbf{V}_1$ undergo layer normalization (LN) and multi-attention (MA) to extract features at different levels, resulting in the within-block features $\mathbf{E}_F \in \mathbb{R}^{N \times D}$.

Next, we transform $\mathbf{E}_F$, $\mathbf{E}_A$, and $\mathbf{E}_{\text{enc}}$ into within-block representations $\mathbf{E}_F^{Q_2}$, $\mathbf{E}_A^{Q_2}$, $\mathbf{E}_{\text{enc}}^{K_2}$, and $\mathbf{E}_{\text{enc}}^{V_2}$ via MLP layers. The queries, keys, and values for the second LN \& MA layer are obtained using the following functions:
\begin{equation}
    \mathbf{Q}_2 = \mathbf{\beta}_3 \times \mathbf{E}_F^{Q_2} + (1 - \mathbf{\beta}_3) \times \mathbf{E}_A^{Q_2} \nonumber
\end{equation}
\begin{equation}
    \mathbf{K}_2 = \mathbf{E}_{\text{enc}}^{K_2} + \mathbf{PE}
\end{equation}
\begin{equation}
    \mathbf{V}_2 = \mathbf{E}_{\text{enc}}^{V_2} \nonumber
\end{equation}
where $\mathbf{PE}$ represents the sinusoidal positional embedding, and $\mathbf{\beta}_3$ is another learnable parameter.

The output of the second LN \& MA layer is then passed through a feed-forward network (FFN), producing the next decoder embedding. Finally, the output embedding $\mathbf{E}_{\text{dec}}$ is sent to the output layer, which consists of a linear layer followed by a sigmoid activation function to predict target-speaker voice activities. The output layer's structure also determines the length of the decoding process.

\subsection{MA-MSE with Deep Interactive Module}
\label{IDM}

The memory-aware multi-speaker embedding (MA-MSE) module is designed to retrieve clean and discriminative multi-speaker embeddings from memory using a simple additive attention mechanism. As outlined in \cite{MAMSE}, the core of the MA-MSE module is the memory component, which consists of speaker embedding basis vectors derived from additional datasets. Specifically, these embedding basis vectors are obtained by clustering speaker embeddings (e.g., i-vectors or x-vectors) and selecting the cluster centers.

Before feeding the features $\mathbf{F'}$ into the MA-MSE module, we apply a clustering-based approach to obtain a speaker activity mask $\mathbf{S} \in \mathbb{R}^{N \times T}$, where each frame is assigned a $0/1$ label indicating speaker presence. The features $\mathbf{F'}$ and the mask $\mathbf{S}$ are then multiplied to select the relevant features for each speaker, yielding the selected features $\mathbf{F}_S = \left[\mathbf{F}_S^1, \mathbf{F}_S^2, \dots, \mathbf{F}_S^N\right]^T \in \mathbb{R}^{N \times D}$. An additive attention mechanism is then employed to match the current speech segment with the most relevant speaker embedding bases from the memory.
\begin{figure}[htbp]
    \centering
    \includegraphics[width=0.5\textwidth]{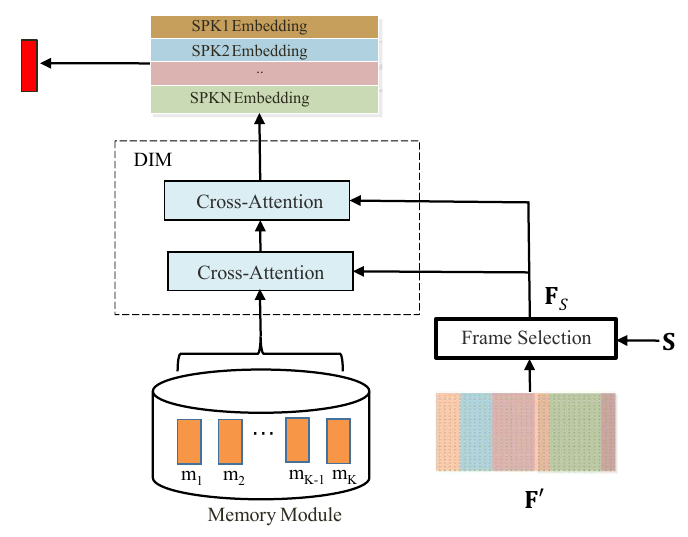}
    \caption{Deep interactive module}
    \label{fig:DIM}
\end{figure}
Through the CHiME-7 DASR Challenge, we identified that if the MA-MSE module structure is not optimized, it can severely affect performance in complex acoustic environments. Furthermore, overly simplistic mechanisms may limit the potential for performance improvement. To address this, we introduce the Deep Interactive Module (DIM), which replaces the additive attention mechanism with a dot-product attention mechanism and increases the depth of interaction layers. This multi-scale feature fusion approach enhances the extraction of cleaner, more discriminative multi-speaker embeddings from the memory module.

The DIM consists of three DIM blocks, each containing two cross-attention layers along the feature dimension. The speaker embedding basis vectors in the memory module are denoted by $\mathbf{M} \in \mathbb{R}^{K \times D_M}$, where $K$ is the number of vectors. In the first DIM block, the input features $\mathbf{F}_S^n$ of the $n$-th speaker and the memory $\mathbf{M}$ are processed as follows:
\begin{equation}
    \mathbf{H}_1^n = \text{Softmax}\left(\frac{\left(\mathbf{F}_S^n\mathbf{W}_1^{n,q}\right)\left({\mathbf{M}\mathbf{W}_1^{n,k}}\right)^T}{\sqrt{d_m}}\right) \mathbf{M}
\end{equation}
where $\mathbf{W}_1^{n,q} \in  \mathbb{R}^{D \times D}$ and $\mathbf{W}_1^{n,k} \in  \mathbb{R}^{D_M \times D}$ are learnable weight matrices, and $\sqrt{d_m}$ is used for scaling to ensure numerical stability. The output of the first DIM block is then calculated by:
\begin{equation}
    \mathbf{H}_2^n = \text{Softmax}\left(\frac{\left(\mathbf{F}_S^n\mathbf{W}_2^{n,q}\right)\left({\mathbf{H}_1^n\mathbf{W}_2^{n,k}}\right)^T}{\sqrt{d_m}}\right) \mathbf{H}_1^n
\end{equation}
where $\mathbf{W}_2^{n,q} \in  \mathbb{R}^{D \times D}$ and $\mathbf{W}_2^{n,k} \in  \mathbb{R}^{D_M \times D}$ are additional learnable weights. After this, the resulting $\mathbf{H}_2^n$ is passed, along with $\mathbf{F}_S^n$, to the next DIM block.

Finally, after processing through all three DIM blocks, the MA-MSE embedding $\mathbf{E}_M$ is obtained. This embedding, which provides crucial supplementary speaker information, is concatenated with the current speaker's i-vector to generate the aggregate embedding $\mathbf{E}_A$.

\subsection{Shared and Soft Mixture of Experts}

The SS-MoE module consists of a shared expert, multiple collaborative experts, input slots, an input dispatch module, and an output combination module.

We denote the input tokens of a sequence as \(\mathbf{X} \in \mathbb{R}^{m \times d}\), where \(m\) is the number of tokens and \(d\) is their dimensionality. The shared expert directly processes the input tokens and produces:

\begin{equation}
    \mathbf{Y}_{\text{sha}}  = \text{Expert}_{\text{sha}}(\mathbf{X}) \in \mathbb{R}^{m \times d}
\end{equation}

In the SS-MoE module, the input dispatch module assigns weights to input tokens and distributes them to different slots. This process ensures that each collaborative expert receives a weighted average of tokens as input, rather than individual tokens. After processing their inputs, the collaborative experts’ outputs are merged by the output combination module, resulting in the fused expert output \(\mathbf{Y}_{\text{co}}\). Each output token is also a weighted average of all collaborative expert outputs.

Finally, the fused expert output and the shared expert output are combined to produce the final output of the SS-MoE module:

\begin{equation}
    \mathbf{Y} = \mathbf{Y}_{\text{sha}} + \mathbf{Y}_{\text{co}}  \in \mathbb{R}^{m \times d}
\end{equation}

Next, we elaborate on the technical details of each component.

\subsubsection{Input Dispatch Module}
\label{sec:input_module}
\begin{figure}[htbp]
    \centering
    \includegraphics[width=0.3\textwidth]{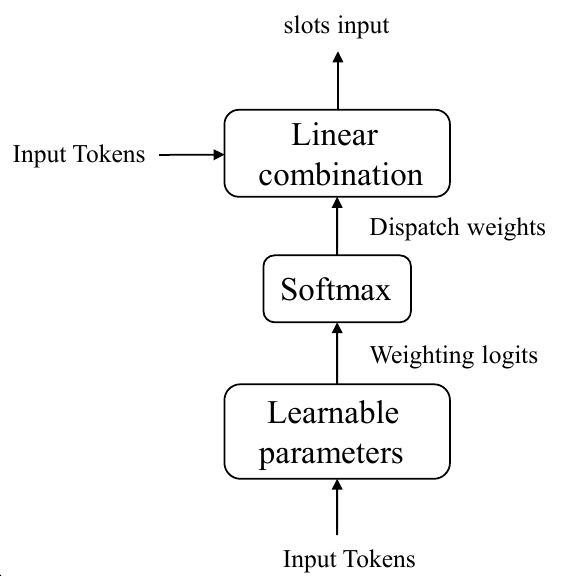}
    \caption{Illustration of the input dispatch module}
    \label{fig:dispatch}
\end{figure}

The architecture of the input dispatch module is illustrated in Figure~\ref{fig:dispatch}. Each SS-MoE layer contains \(n\) collaborative experts. Each expert processes \(p\) slots, and each slot is associated with a learnable \(d\)-dimensional vector \(\Phi \in \mathbb{R}^{d \times (n \cdot p)}\). The weight logits are computed as the matrix product of \(\mathbf{X}\) and \(\Phi\):

\begin{equation}
    \text{Weight\_logits} = (\mathbf{X} \Phi) \in \mathbb{R}^{m \times (n \cdot p)}
\end{equation}

Then, a Softmax is applied to each column of the weight logits:

\begin{equation}
    \mathbf{D}_{ij} = \frac{\exp(\text{Weight\_logits}_{ij})}{\sum_{i'=1}^{m} \exp(\text{Weight\_logits}_{i'j})} \in \mathbb{R}^{m \times (n \cdot p)}
\end{equation}

Here, \(\mathbf{D}_{ij}\) is referred to as the dispatch weight. The inputs are then linearly combined based on these weights to obtain the inputs \(\tilde{\mathbf{X}}\) for each of the \(p\) slots of the \(n\) collaborative experts:

\begin{equation}
    \tilde{\mathbf{X}} = \mathbf{D}^{\top} \mathbf{X} \in  \mathbb{R}^{(n \cdot p) \times d}
\end{equation}

Intuitively, each slot in \(\tilde{\mathbf{X}}\) represents a weighted sum of all input tokens in \(\mathbf{X}\).

\subsubsection{Output Combined Module}
\begin{figure}[htbp]
    \centering
    \includegraphics[width=0.3\textwidth]{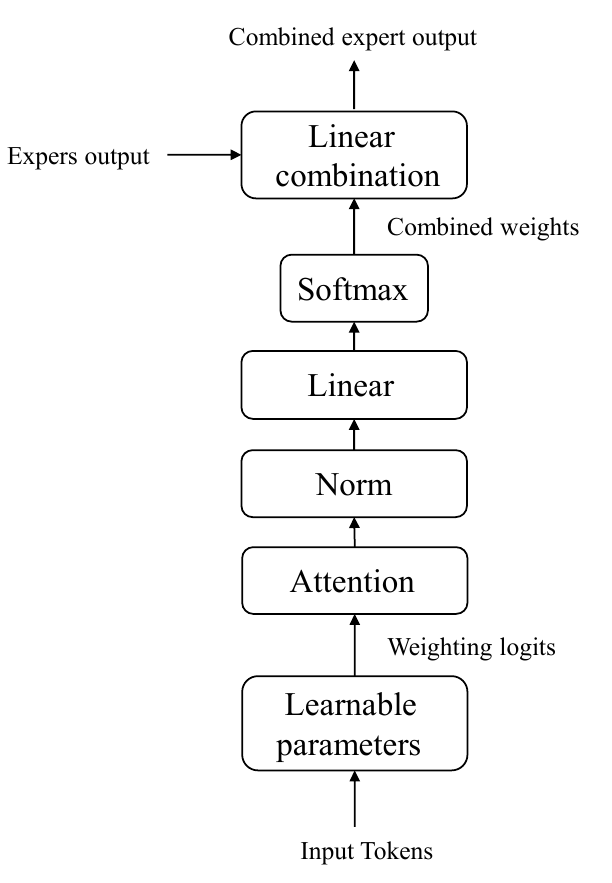}
    \caption{Illustration of the output combined module}
    \label{fig:combine}
\end{figure}

The purpose of the output combination module is to better fuse the outputs of multiple experts. Its architecture is shown in Figure~\ref{fig:combine}. The outputs of the collaborative experts are defined as:

\begin{equation}
    \tilde{Y}_{\text{co}} = \text{Experts}_{\text{co}}(\tilde{\mathbf{X}}) \in \mathbb{R}^{(n \cdot p) \times d}
\end{equation}

The output combination module further transforms the weight logits from Section~\ref{sec:input_module}. Specifically, an attention layer is used to focus on the most informative weights:

\begin{equation}
    \begin{aligned}
    Q &= \text{Weight\_logits} \cdot W_Q \\
    K &= \text{Weight\_logits} \cdot W_K \\
    V &= \text{Weight\_logits} \cdot W_V \\
    \text{Logits\_attention} &= \text{Softmax}\left(\frac{QK^\top}{\sqrt{d_k}}\right) V \in \mathbb{R}^{m \times (n \cdot p)}
    \end{aligned}
\end{equation}

Here, \( W_Q, W_K, W_V \in \mathbb{R}^{(n \cdot p) \times (n \cdot p)} \) are learnable parameters. The attention output is then normalized:

\begin{equation}
    \text{Logits\_norm} = \text{Norm}(\text{Logits\_attention})  \in \mathbb{R}^{m \times (n \cdot p)}
\end{equation}

We use Instance Normalization for this normalization. A linear layer is then applied to project the normalized logits into a suitable feature space, yielding the combined logits:

\begin{equation}
    \text{Combined\_logits} = \text{Linear}(\text{Logits\_norm})  \in \mathbb{R}^{m \times (n \cdot p)}
\end{equation}

A row-wise Softmax is applied to the combined logits:

\begin{equation}
\mathbf{C}_{ij} = \frac{\exp(\text{Combined\_logits}_{ij})}{\sum_{j'=1}^{n \cdot p} \exp(\text{Combined\_logits}_{ij'})} \in \mathbb{R}^{m \times (n \cdot p)}
\end{equation}

\(\mathbf{C}_{ij}\) is referred to as the combination weight. Finally, these weights are used to linearly combine the outputs \(\tilde{Y}_{\text{co}}\) from all collaborative experts to produce the fused expert output:

\begin{equation}
    \mathbf{Y}_{\text{co}} = \mathbf{C} \tilde{Y}_{\text{co}} \in \mathbb{R}^{m \times d}
\end{equation}

\subsection{Parameter Transfer Strategy for Accelerating Model Optimization}
Training a Mixture-of-Experts (MoE) model from scratch typically incurs substantial computational and time costs. To address this challenge, we propose a parameter transfer strategy aimed at leveraging pretrained model parameters to effectively initialize the NSD-MS2S-SSMoE model. This approach enables faster convergence and reduces training overhead.

\begin{figure}[htbp]
    \centering
    \includegraphics[width=0.5\textwidth]{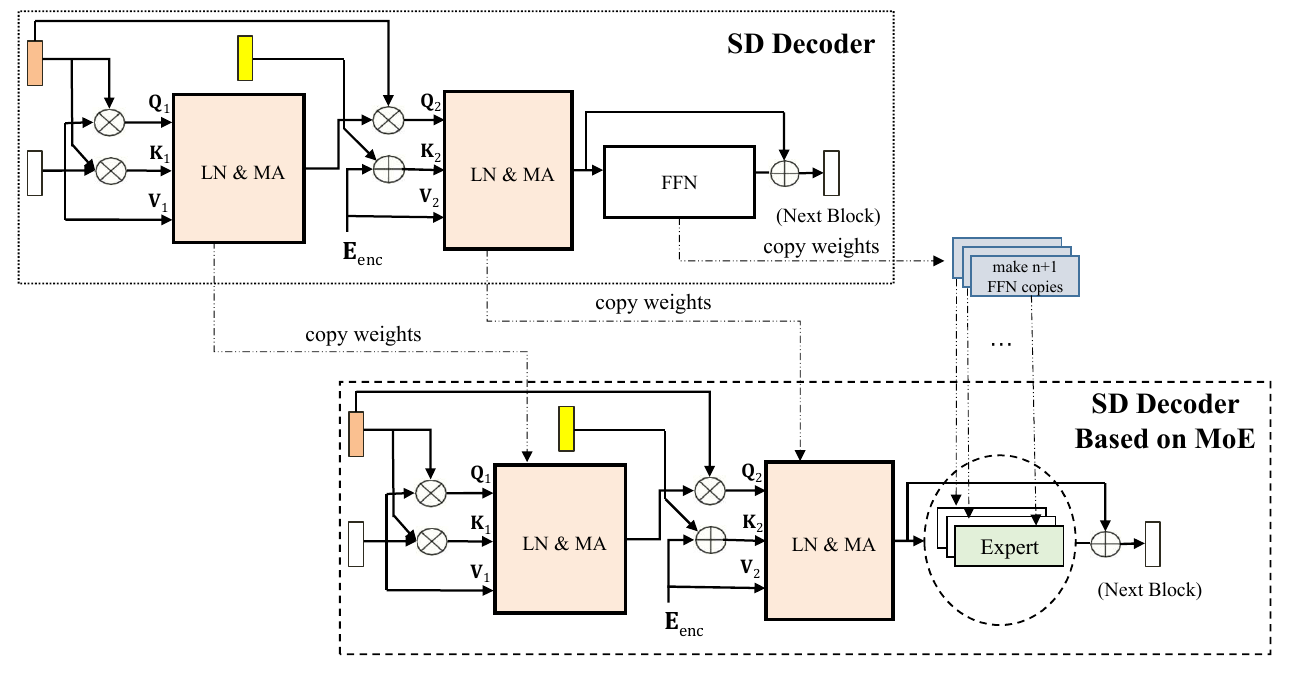}
    \caption{Illustration of the parameter transfer process from the pretrained NSD-MS2S model to the NSD-MS2S-SSMoE model.}
    \label{fig:cw}
\end{figure}

Specifically, we utilize a pretrained NSD-MS2S model and transfer its parameters to initialize structurally compatible components within the NSD-MS2S-SSMoE model. As illustrated in Figure~\ref{fig:cw}, we identify all modules that are shared between the two models and perform direct parameter copying wherever applicable.

In particular, for the speaker decoder block, we replicate the Feed-Forward Network (FFN) parameters from the pretrained model \(n + 1\) times to initialize the \(n + 1\) expert networks in the SSMoE module. Other identical submodules, such as attention mechanisms and normalization layers, are directly initialized with their corresponding pretrained weights.

This parameter reuse paradigm allows us to retain the inductive biases and prior knowledge embedded in the pretrained model, thereby providing a strong initialization for the SSMoE-enhanced architecture. Empirical results demonstrate that this transfer strategy significantly reduces training cost while preserving or even improving model performance.
\section{Experimental Setup}
\subsection{Experimental Setup}
\subsubsection{Datasets}
To evaluate the robustness of the proposed diarization system in complex acoustic conditions, experiments were conducted on three challenging English datasets: CHiME-6, DiPCo, and Mixer 6. Additionally, we further validated the proposed method on the DIHARD-III dataset, which includes a broader range of scenarios.

\subsubsection{Training Data}

For the CHiME-6, DiPCo, and Mixer 6 datasets, we adopt a simulation strategy\footnote{\url{https://github.com/jsalt2020-asrdiar/jsalt2020\_simulate}} to generate a large amount of synthetic data for training set augmentation. Specifically, simulated multi-speaker conversations are constructed using real single-speaker utterances. This approach enables the expansion of the training data without incurring the cost of manual annotation. In addition to the official training sets of CHiME-6 and Mixer 6, we further enhance the Mixer 6 training set by applying pseudo-labeling techniques, following the method proposed in \cite{10389615}. The total duration of training data used amounted to approximately 5,300 hours. 

Since the DIHARD-III dataset does not provide a dedicated training set, it poses a significant challenge to the generalization capability of the diarization system. To address this, in addition to simulating multi-speaker conversations using the LibriSpeech dataset, we utilized several real-world datasets, including Switchboard-1~\cite{king2005svitchboard}, AMI~\cite{kraaij2005ami}, and the development portion of VoxConverse~\cite{chung2020spot}. The total duration of training data used amounted to approximately 1,400 hours.

\subsubsection{Initialization of Diarization Systems}
For the CHiME-6, DiPCo, and Mixer 6 datasets, we adopt the baseline VAD model provided by CHiME-7\footnote{\url{https://github.com/espnet/espnet/tree/master/egs2/chime7_task1/diar_asr1}}, and further fine-tune it using the training data from CHiME-6 and Mixer 6. After applying VAD, we segment the detected speech regions into overlapping subsegments with a window length of 1.5 seconds and a shift of 0.75 seconds. We then extract x-vectors from each segment using the ECAPA-TDNN model \cite{desplanques2020ecapa}, which is pretrained on the VoxCeleb corpus \cite{chung2018voxceleb2}. Finally, speaker clustering is performed using spectral clustering based on cosine similarity.

For the DIHARD-III dataset, a clustering-based diarization system, VBx~\cite{landini2022bayesian}, was adopted for initialization. Specifically, silence segments were first removed based on official annotations. Then, x-vectors were extracted using a speaker embedding model (ECAPA-TDNN) pre-trained on VoxCeleb. Agglomerative Hierarchical Clustering (AHC) was performed on the x-vectors to obtain coarse cluster assignments, which were used to initialize the parameters of VBx. In this system, each state in the Hidden Markov Model (HMM) is treated as an individual speaker, and transitions between states correspond to speaker changes. The x-vector sequence is regarded as the observation sequence, and Variational Inference is employed to obtain the most probable state sequence, corresponding to the final diarization output.

\subsubsection{Model Configuration and Training}
For the NSD-MS2S-MoE system, 40-dimensional log Mel-filterbank (Fbank) features are used as input. The model consists of 6 speaker detection encoder layers and 6 speaker detection decoder layers. For the CHiME-6, DiPCo, and Mixer 6 datasets, the SS-MoE module was inserted into the last three decoder layers. For the DIHARD-III dataset, SS-MoE was only applied to the second decoder layer.

Each SS-MoE block comprises Gated Linear Unit (GLU)-based expert models, each consisting of two fully-connected layers. The first layer projects the input to $2d$ dimensions, followed by Gaussian Error Linear Unit (GEGLU) gating. The input is split along the channel dimension, where one half is passed through a GELU activation function and multiplied element-wise with the other half. Dropout with a rate of 0.1 is applied for regularization. The second fully-connected layer projects the output back to the original $d$ dimensions, with $d=512$. Each SS-MoE layer contains 6 experts, and the number of input slots is set to 4. In the fusion branches, attention layers use a dimensionality of 512 with 4 attention heads. Other model parameters remain consistent with the baseline NSD-MS2S configuration.

For experiments on the DIHARD-III dataset, we first pretrained the NSD-MS2S model for 30 epochs using a learning rate of 1e-4. The resulting parameters were then used to initialize the corresponding modules of the NSD-MS2S-SSMoE model. During fine-tuning, a two-stage strategy was adopted: first, the learning rate was set to 1e-5 and all parameters except the SS-MoE layers were frozen for 2 epochs; then, all parameters were unfrozen and the entire model was fine-tuned for an additional 3 epochs. The Adam optimizer was used throughout. This staged fine-tuning approach facilitates stable training while gradually improving the SS-MoE performance.

For the CHiME-6, DiPCo, and Mixer 6 datasets, pretraining was performed for 6 epochs. During fine-tuning, the model was first trained for 1 epoch with frozen parameters except for the SS-MoE layers, followed by another epoch with all parameters unfrozen. Other experimental settings remained the same.

\subsubsection{Baseline Systems}
To assess the effectiveness of the proposed NSD-MS2S-MoE system, we compared it against several state-of-the-art diarization systems, including TS-VAD, NSD-MA-MSE, and NSD-MS2S. Furthermore, we included QM-TS-VAD~\cite{niu2023unsupervised}, a recent TS-VAD variant, and ITS-VAD, the top-ranked system in the DIHARD-III challenge, as additional baselines.

\subsubsection{Evaluation Metrics}
For the DIHARD-III dataset, the Diarization Error Rate (DER) was used as the primary evaluation metric, with a collar tolerance of 0 seconds to ensure strict alignment with reference annotations. For CHiME-6, DiPCo, and Mixer 6, both DER and the Jaccard Error Rate (JER) were adopted, using a collar of 0.25 seconds. For methods or datasets where reference results were unavailable in the literature, missing results are indicated with “--”.

\section{Results and Analysis}
\subsubsection{Results on Different Datasets}
\setlength{\tabcolsep}{12pt} 
\begin{table}
    \centering
    \setlength{\abovecaptionskip}{0pt}%
    \setlength{\belowcaptionskip}{10pt}%
    \caption{Performance comparison on the CHiME-6 dataset (collar = 0.25\,s)}  
    \label{tab:chime6_moe_results}
    \begin{tabular}{l l S[table-format=2.2] S[table-format=2.2]}  
    \toprule
    \textbf{Method} & \textbf{Set} & \textbf{DER} & \textbf{JER} \\  
    \midrule
    x-vectors + SC & DEV  & 40.32 & 42.31 \\  
    & EVAL & 36.32 & 43.39 \\  
    \cmidrule(r){2-4}  
    DiaPer & DEV  & {-} & {-} \\  
    & EVAL & 69.94 & {-} \\  
    \cmidrule(r){2-4}  
    EEND-VC\cite{kamo2024ntt} & DEV  &  36.60 &  {-} \\
    &  EVAL &  39.80 &  {-} \\
    \cmidrule(r){2-4}  
    TS-VAD & DEV  & 33.79 & 35.16 \\  
    & EVAL & 32.45 & 38.12 \\  
    \cmidrule(r){2-4}  
    NSD-MA-MSE & DEV  & 32.27 & 34.76 \\  
    & EVAL & 32.09 & 37.61 \\  
    \cmidrule(r){1-4}  
    NSD-MS2S  & DEV  & 28.36 & 31.49 \\  
    & EVAL & 29.45 & 33.84 \\  
    \cmidrule(r){2-4} 
    NSD-MS2S-SoftMoE  & DEV  & 27.53 & 30.63 \\  
    & EVAL & 29.12 & 33.21 \\  
    \cmidrule(r){2-4} 
    NSD-MS2S-SSMoE (ours)  & DEV  & \textbf{26.31} & \textbf{28.56} \\  
    & EVAL & \textbf{28.51} & \textbf{32.31} \\  
    \bottomrule  
    \end{tabular}
\end{table}

Table~\ref{tab:chime6_moe_results} reports the diarization performance of different systems on the CHiME-6 dataset. The proposed NSD-MS2S-SSMoE achieves the lowest DER and JER on both the development and evaluation sets, with DER/JER values of 26.31\%/28.56\% and 28.51\%/32.31\%, respectively. Compared with the NSD-MS2S baseline, the DER is relatively reduced by 7.23\% on the development set and 3.19\% on the evaluation set, demonstrating the effectiveness of incorporating sparse mixture-of-experts in improving diarization accuracy.

\begin{table}
    \centering
    \caption{Performance comparison on the DiPCo dataset (collar = 0.25\,s).}
    \label{tab:dipco_moe_results}
    \begin{tabular}{l l S[table-format=2.2] S[table-format=2.2]}
    \toprule
    \textbf{Method} & \textbf{Split} & \textbf{DER} & \textbf{JER} \\
    \midrule
    x-vectors + Spectral Clustering & DEV & 24.47 & 28.97 \\
    & EVAL & 25.18 & 35.08 \\
    \cmidrule(r){2-4}
    DiaPer & DEV & {-} & {-} \\
    & EVAL & 43.37 & {-} \\
    \cmidrule(r){2-4}
    EEND-VC~\cite{kamo2024ntt} & DEV & 22.80 & {-} \\
    & EVAL & 26.90 & {-} \\
    \cmidrule(r){2-4}
    TS-VAD & DEV & 22.14 & 25.33 \\
    & EVAL & 23.45 & 33.26 \\
    \cmidrule(r){2-4}
    NSD-MA-MSE & DEV & 21.04 & 24.01 \\
    & EVAL & 22.78 & 31.34 \\
    \cmidrule(r){1-4}
    NSD-MS2S & DEV & 17.06 & 18.54 \\
    & EVAL & 19.31 & 26.63 \\
    \cmidrule(r){2-4}
    NSD-MS2S-SoftMoE & DEV & 16.87 & 18.37 \\
    & EVAL & 19.42 & 26.89 \\
    \cmidrule(r){2-4}
    NSD-MS2S-SSMoE (Ours) & DEV & \textbf{15.97} & \textbf{17.17} \\
    & EVAL & \textbf{19.25} & \textbf{26.25} \\
    \bottomrule
    \end{tabular}
\end{table}

Table~\ref{tab:dipco_moe_results} presents results on the DiPCo dataset. Our NSD-MS2S-SSMoE consistently outperforms all baselines. In particular, DER/JER are reduced to 15.97\%/17.17\% on the development set and 19.25\%/26.25\% on the evaluation set. Compared to the NSD-MS2S baseline, relative DER reduction on the development set reaches 6.39\%, while the gain on the evaluation set is more modest (1.09\%), indicating a possible risk of overfitting due to increased model complexity in the expert architecture.

\begin{table}
    \centering
    \caption{Performance comparison on the Mixer 6 dataset (collar = 0.25\,s).}
    \label{tab:mixer6_moe_results}
    \begin{tabular}{l l S[table-format=2.2] S[table-format=2.2]}
    \toprule
    \textbf{Method} & \textbf{Split} & \textbf{DER} & \textbf{JER} \\
    \midrule
    x-vectors + Spectral Clustering & DEV & 15.80 & 23.07 \\
    & EVAL & 9.53 & 12.08 \\
    \cmidrule(r){2-4}
    DiaPer & DEV & {-} & {-} \\
    & EVAL & 10.99 & {-} \\
    \cmidrule(r){2-4}
    EEND-VC~\cite{kamo2024ntt} & DEV & 10.30 & {-} \\
    & EVAL & 6.80 & {-} \\
    \cmidrule(r){2-4}
    TS-VAD & DEV & 9.67 & 13.79 \\
    & EVAL & 6.18 & 7.10 \\
    \cmidrule(r){2-4}
    NSD-MA-MSE & DEV & 9.28 & 12.94 \\
    & EVAL & 6.21 & 7.12 \\
    \cmidrule(r){1-4}
    NSD-MS2S & DEV & 7.27 & 9.56 \\
    & EVAL & 5.01 & 5.54 \\
    \cmidrule(r){2-4}
    NSD-MS2S-SoftMoE & DEV & 7.25 & 9.38 \\
    & EVAL & 4.99 & 5.52 \\
    \cmidrule(r){2-4}
    NSD-MS2S-SSMoE (Ours) & DEV & \textbf{7.16} & \textbf{9.14} \\
    & EVAL & \textbf{4.94} & \textbf{5.49} \\
    \bottomrule
    \end{tabular}
\end{table}

Table~\ref{tab:mixer6_moe_results} summarizes performance on the Mixer 6 dataset. NSD-MS2S-SSMoE achieves the best DER and JER in both splits, with DER/JER of 7.16\%/9.14\% on the development set and 4.94\%/5.49\% on the evaluation set. However, since most systems already achieve very low error rates on this dataset and potential annotation inaccuracies may limit further improvement, the performance gains here are marginal.

\setlength{\tabcolsep}{1pt} 
\begin{table*}[htbp]
    \centering
    \caption{DER (\%) comparison of different systems on the DIHARD-III evaluation set across eight domains (collar = 0s).}
    \label{tab:dihard3_der}
    \resizebox{\textwidth}{!}{ 
    \begin{tabular}{l|c|c|c|c|c|c|c|cc}
    \hline
    \textbf{System} & \textbf{BROADC.} & \textbf{COURT} & \textbf{MAP TASK} & \textbf{CLINICAL} & \textbf{SOC.LAB} & \textbf{SOC.FIELD} & \textbf{CTS} & \textbf{MEETING} \\ \hline
    VBx            & 4.22           & 3.08            & 3.41           & 11.08            & 6.04            & 8.05            & 14.20           & 33.20           \\ 
    TS-VAD         & 5.56           & 3.88            & 4.75           & 15.87            & 6.71            & 9.45            & 6.41            & 31.13           \\ 
    ITS-VAD        & 4.46           & 3.07            & 3.20           & 10.03            & 3.81            & 7.10            & 6.11            & 28.17           \\ 
    QM-TS-VAD\cite{niu2023unsupervised}      & 4.23           & 3.12            & \textbf{1.59}  & \textbf{9.81}    & \textbf{3.69}   & 6.67            & 5.85            & 27.48           \\ 
    NSD-MA-MSE     & 4.29           & 2.85            & 2.64           & 9.96             & 4.98            & 7.45            & 5.78            & 28.41           \\ 
    \cmidrule(r){1-9} 
    NSD-MS2S       & 4.28           & \textbf{2.75}   & 5.40           & 12.28            & 4.01            & 6.59            & \textbf{5.37}   & 26.89           \\ 
    NSD-MS2S-SoftMoE   & 4.28  & 2.81   & 5.32           & 11.98         & 4.12      & 6.48   & 5.54   & 26.77  \\ 
    \textbf{NSD-MS2S-SSMoE (ours)}  & \textbf{4.21}  & \textbf{2.75}   & 5.14           & 11.48     & 3.85            & \textbf{6.24}   & \textbf{5.37}   & \textbf{26.49}  \\ 
    \bottomrule
\end{tabular}
    }
\end{table*}

\noindent
 We present in Table~\ref{tab:dihard3_der} the diarization error rate (DER) results of various systems on the DIHARD-III evaluation set across eight domains: broadcast news (BROADC.), courtroom (COURT), map task (MAP TASK), clinical interviews (CLINICAL), sociolinguistic lab interviews (SOC.LAB), sociolinguistic field recordings (SOC.FIELD), conversational telephone speech (CTS), and meetings (MEETING).

Our proposed system, NSD-MS2S-SSMoE, achieves the SOTA performance in multiple domains including BROADC., COURT, SOC.LAB, SOC.FIELD, CTS, and MEETING, demonstrating strong robustness and generalization across diverse acoustic conditions.

Additionally, QM-TS-VAD shows superior results in MAP TASK and CLINICAL, likely benefiting from its fine-tuning on simulated data generated from high-quality in-domain recordings, which enhances performance in domain-specific settings.

It is also worth noting that in the BROADC. domain, all end-to-end diarization systems struggle to surpass the traditional VBx system. This is likely due to the very low overlap speech ratio (only 1.18\%) in this domain, which limits the advantage of overlap-aware modeling typically offered by end-to-end systems.

\subsubsection{Analysis of the DIM Module Results}
\begin{figure}[htbp]
    \centering
    \includegraphics[width=0.5\textwidth]{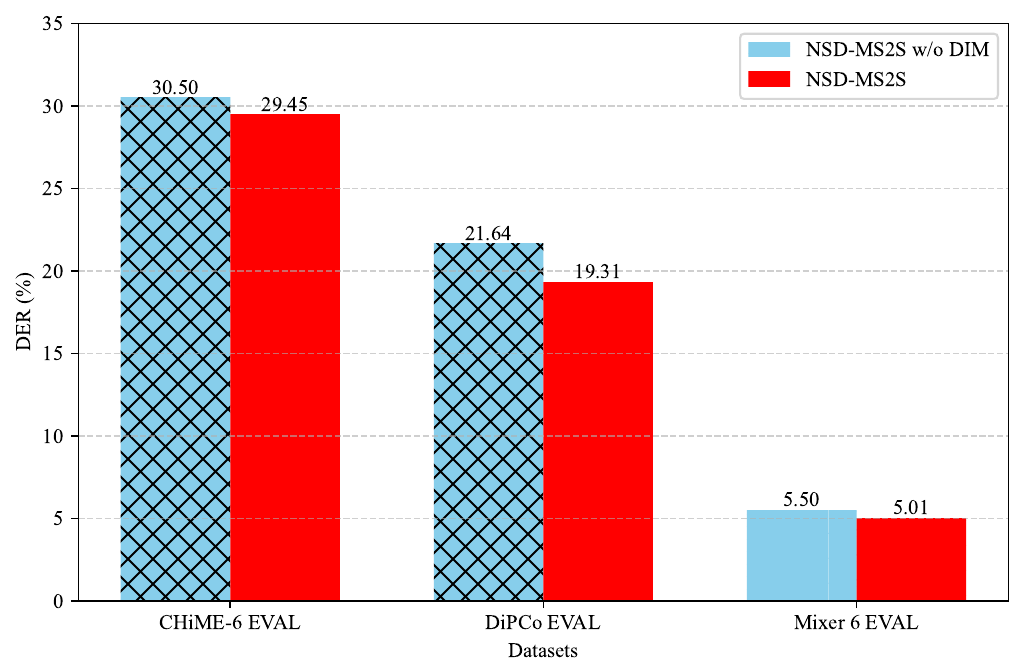}
    \caption{Impact of the DIM module on the performance of the NSD-MS2S system}
    \label{fig:DIM_infulence}
\end{figure}

Figure~\ref{fig:DIM_infulence} illustrates the impact of the proposed DIM module on the performance of the NSD-MS2S system. It can be observed that the inclusion of the DIM module consistently improves system performance across different datasets. Specifically, the DIM module reduces the Diarization Error Rate (DER) on the evaluation sets of CHiME-6, DiPCo, and Mixer 6 by 3.44\% (from 30.50\% to 29.45\%), 10.76\% (from 21.64\% to 19.31\%), and 9.80\% (from 5.50\% to 5.01\%), respectively.

\begin{figure}[htbp]
    \centering
    \includegraphics[width=0.5\textwidth]{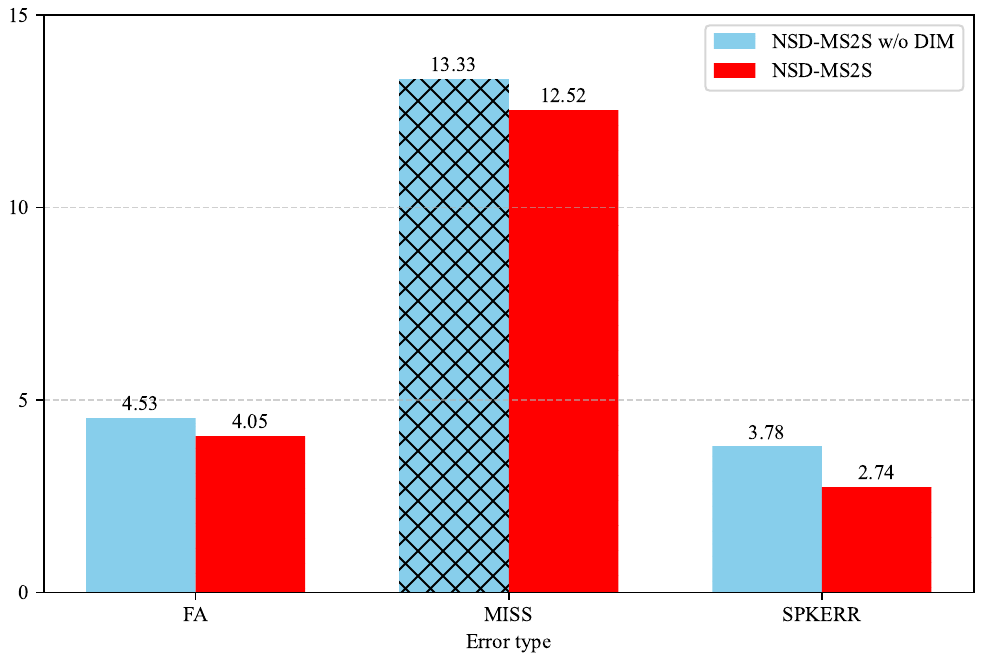}
    \caption{Detailed breakdown of DER components on the DiPCo evaluation set}
    \label{fig:DIM_infulence2}
\end{figure}

To further investigate the effect of the DIM module, we analyze the changes in the components of DER on the DiPCo evaluation set, as shown in Figure~\ref{fig:DIM_infulence2}. The DIM module demonstrates varying degrees of improvement across all error types, including False Alarm (FA), Miss (MISS), and Speaker Error (SPKERR). Notably, the most significant improvement is observed in SPKERR, which is relatively reduced by 27\% (from 3.78\% to 2.74\%). These results indicate that the DIM module helps the NSD-MS2S system extract cleaner and more discriminative speaker embeddings, thereby enhancing its ability to differentiate between speakers effectively.

\subsubsection{Convergence analysis of parameter migration strategies}

\begin{figure}[htbp]
    \centering
    \includegraphics[width=0.5\textwidth]{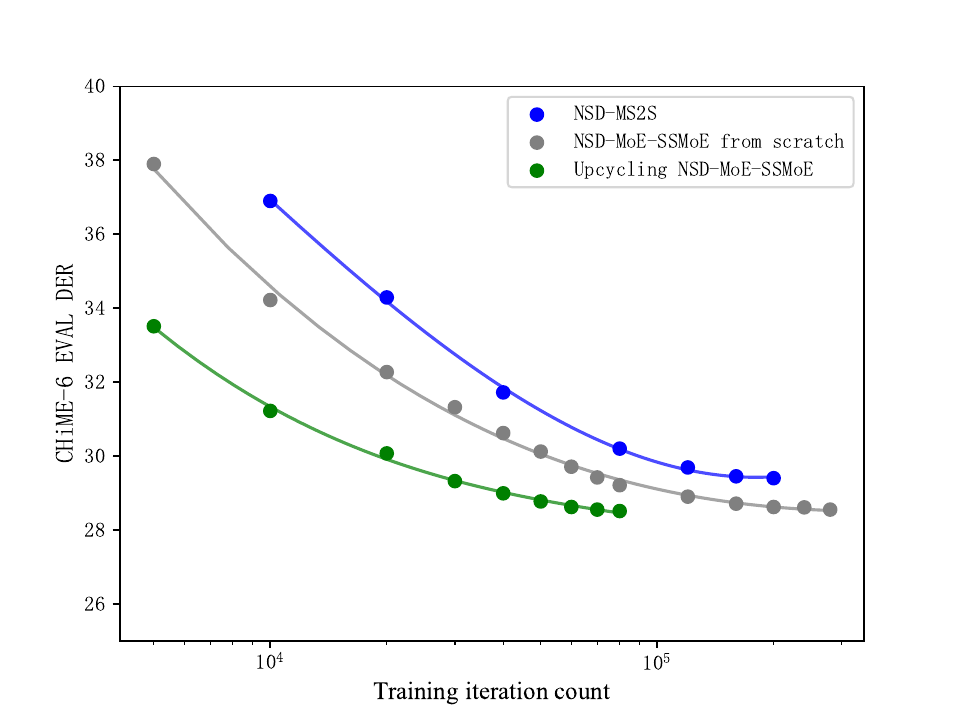}
    \caption{Convergence comparison with different setups.}
    \label{fig:cwe}
\end{figure}

Figure~\ref{fig:cwe} illustrates the convergence behavior of NSD-MS2S and NSD-MS2S-SSMoE on the CHiME-6 evaluation set. The x-axis represents the logarithm of the number of model update iterations, while the y-axis indicates the DER on the CHiME-6 evaluation set. Blue dots correspond to NSD-MS2S, gray dots to NSD-MS2S-SSMoE initialized with random parameters, and green dots to NSD-MS2S-SSMoE initialized with pretrained NSD-MS2S parameters. In the early training stages, the model initialized with random parameters exhibits more volatile updates, likely due to a relatively high learning rate. As training progresses, NSD-MS2S-SSMoE converges to a lower DER compared to NSD-MS2S, which is consistent with the previous experimental results. Furthermore, the model utilizing parameter transfer exhibits smoother convergence and reaches the optimal region more rapidly, reducing the retraining cost by over 50\%.

\subsubsection{Effect of the Number of Experts on System Performance}

\begin{figure}[htbp]
    \centering
    \includegraphics[width=0.5\textwidth]{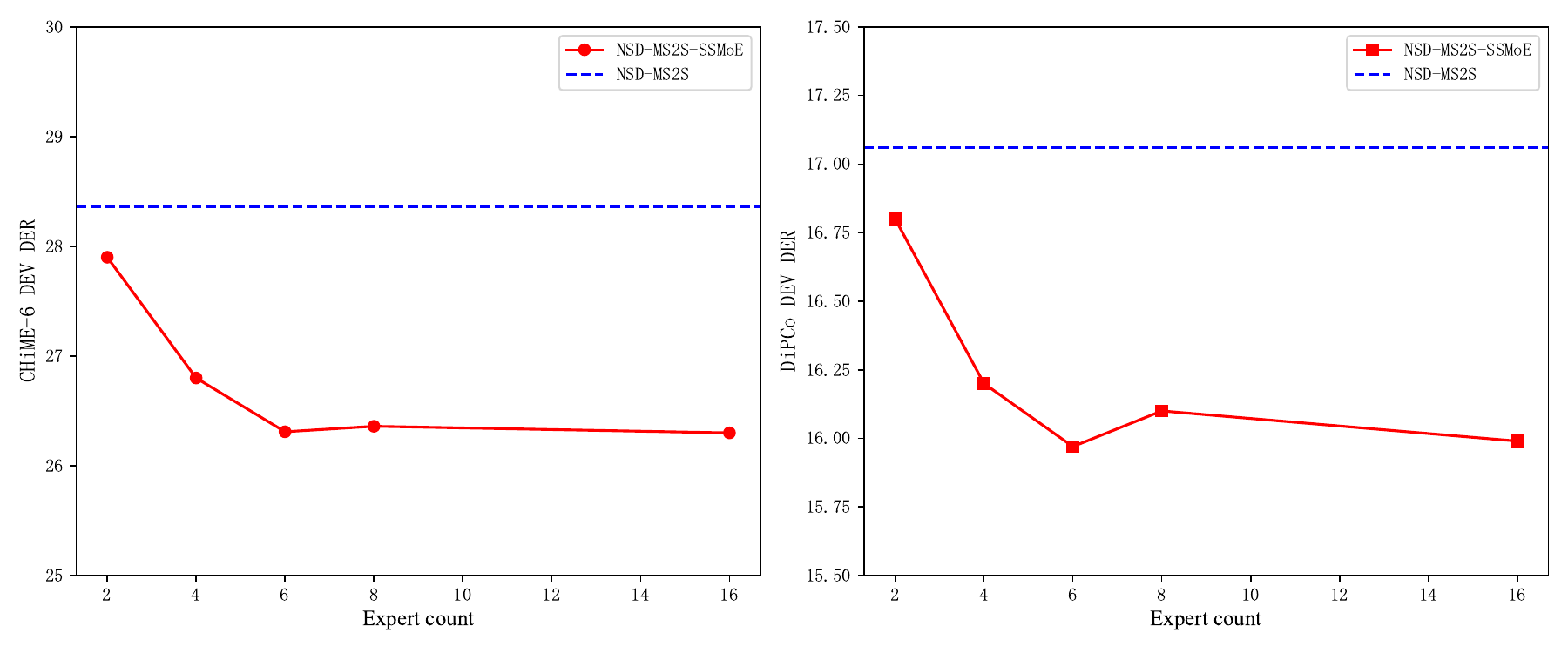}
    \caption{Impact of the number of experts on system performance.}
    \label{fig:expers_ablation}
\end{figure}

Figure~\ref{fig:expers_ablation} shows the impact of the number of experts on system performance, evaluated using DER on the development sets of CHiME-6 and DiPCo. The blue dashed line indicates the baseline performance of NSD-MS2S. As the number of experts increases, the DER of NSD-MS2S-SSMoE decreases, suggesting that incorporating more experts effectively enhances system performance. Specifically, increasing the number of experts from 2 to 6 significantly improves model performance, likely because more experts can better capture the complex speaker characteristics in diverse acoustic scenarios. However, further increasing the number of experts yields marginal gains, indicating performance saturation.

\subsubsection{Effect of Expert Placement on System Performance}

\begin{figure}[htbp]
    \centering
    \includegraphics[width=0.5\textwidth]{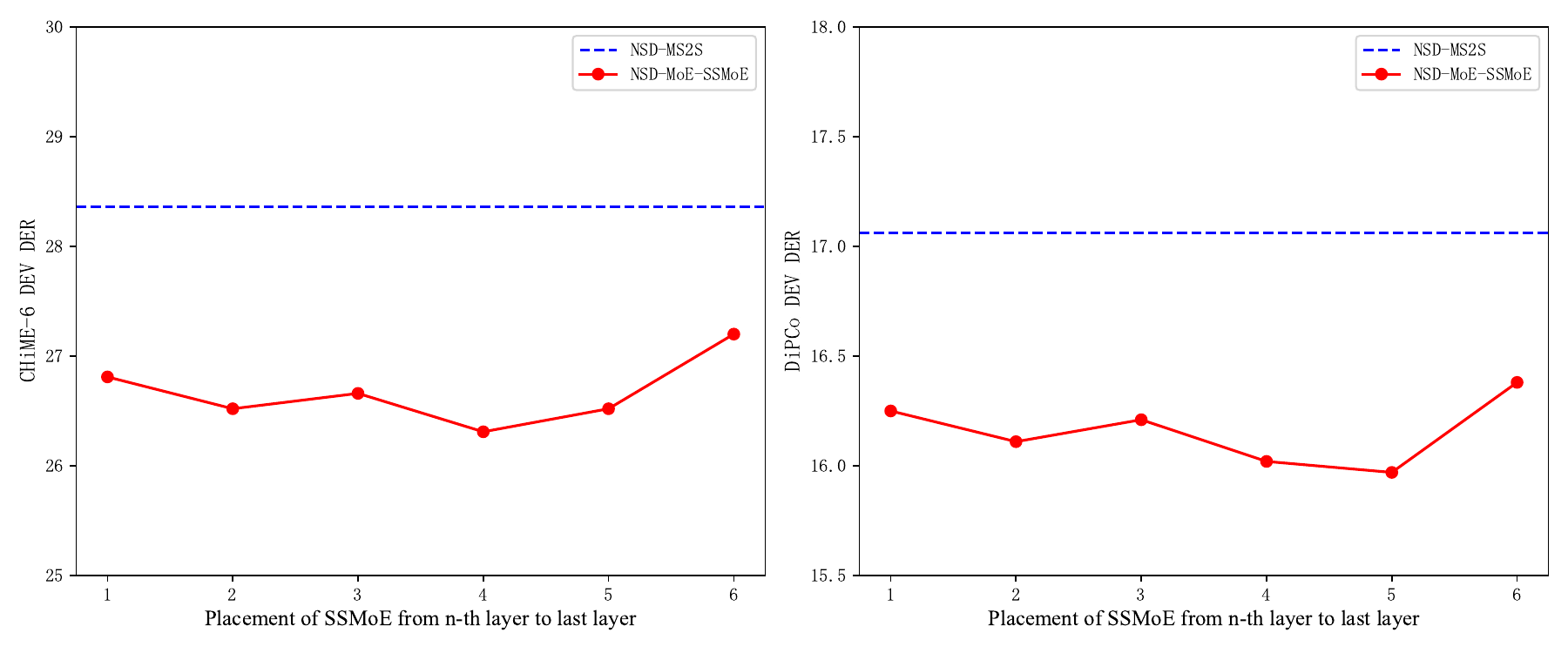}
    \caption{Impact of expert placement on system performance.}
    \label{fig:MoE_place}
\end{figure}

Figure~\ref{fig:MoE_place} presents the impact of expert placement on system performance. Each data point corresponds to inserting SS-MoE modules from the $n$-th layer to the final layer of the speaker decoder. The results indicate that adding MoE modules across all layers does not necessarily yield optimal performance. On the CHiME-6 development set, the best results are obtained when SS-MoE modules are inserted in the last three layers (layers 4–6), while for DiPCo, inserting them in the last two layers (layers 5–6) leads to better performance. These findings suggest that optimal expert placement should be determined on a task-specific basis.

\subsubsection{Comparison Between NSD-MS2S-SSMoE and NSD-MS2S Fusion Models}

Table~\ref{tab:chime_fusion_moe_results} compares the performance of the proposed mixture of experts (MoE) model and a model ensemble approach across three datasets. Here, NSD-MS2S (Fusion) refers to an ensemble-based method where six model checkpoints from different epochs are averaged at the parameter level, effectively mitigating model bias through ensembling. The results show that NSD-MS2S (Fusion) achieves significantly lower DER than the single NSD-MS2S model, highlighting the benefits of ensemble learning.

\begin{table}[htbp]
    \centering
    \caption{Performance comparison of NSD-MS2S-SSMoE and NSD-MS2S fusion models on CHiME-6, DiPCo, and Mixer 6 datasets (collar = 0.25s). AVG denotes the average DER/JER over the three datasets.}
    \renewcommand{\arraystretch}{1.2}
    \setlength{\tabcolsep}{1.pt}
    \begin{tabular}{l l  c c  c c  c c  c c}
        \toprule
        \multicolumn{1}{c}{\multirow{2}{*}{\textbf{Method}}} & \multicolumn{1}{c}{\multirow{2}{*}{\textbf{Set}}} 
        & \multicolumn{2}{c}{\textbf{CHiME-6}} & \multicolumn{2}{c}{\textbf{DiPCo}} 
        & \multicolumn{2}{c}{\textbf{Mixer 6}} & \multicolumn{2}{c}{\textbf{AVG}} \\
        \cmidrule{3-10}
        & & \textbf{DER} & \textbf{JER} & \textbf{DER} & \textbf{JER} & \textbf{DER} & \textbf{JER} & \textbf{DER} & \textbf{JER} \\
        \midrule
        \multirow{2}{*}{NSD-MS2S} 
        & DEV   & 28.36 & 31.49 & 17.06 & 18.54 & 7.27  & 9.56  & 17.56 & 19.86 \\
        & EVAL  & 29.45 & 33.84 & 19.31 & 26.63 & 5.01  & 5.54  & 17.92 & 22.00 \\
        \cmidrule{2-10}
        \multirow{2}{*}{NSD-MS2S (Fusion)} 
        & DEV   & 26.78 & 29.45 & 16.13 & 18.37 & 7.22  & 9.95  & 16.71 & 18.76 \\
        & EVAL  & 28.51 & 32.63 & \textbf{18.83} & \textbf{25.72} & 4.95  & \textbf{5.45}  & \textbf{17.43} & \textbf{21.26} \\
        \cmidrule{2-10}
        \multirow{2}{*}{NSD-MS2S-SSMoE} 
        & DEV   & \textbf{26.31} & \textbf{28.56} & \textbf{15.97} & \textbf{17.17} & \textbf{7.16}  & \textbf{9.41}  & \textbf{16.48} & \textbf{18.38} \\
        & EVAL  & \textbf{28.51} & \textbf{32.31} & 19.25 & 26.25 & \textbf{4.94}  & 5.49  & 17.56 & 21.35 \\
        \bottomrule
    \end{tabular}
    \label{tab:chime_fusion_moe_results}
\end{table}

Furthermore, the NSD-MS2S-SSMoE model outperforms the fusion model on most metrics. However, while NSD-MS2S-SSMoE achieves a lower average DER on the development sets (from 16.71\% to 16.48\%), it slightly underperforms the fusion model on the evaluation sets (17.51\% vs. 17.43\%). This indicates that despite the strong learning capacity of SS-MoE and its effectiveness in alleviating bias, it may still be prone to overfitting, warranting further investigation.

\section{Conclusion}
This paper presents a novel speaker diarization system, NSD-MS2S, which integrates a memory-aware multi-speaker embedding module with a Seq2Seq architecture. By separately modeling acoustic features and speaker embeddings and introducing a lightweight feature fusion method, NSD-MS2S achieves notable improvements in both efficiency and performance, especially in challenging scenarios with high noise, heavy overlap, and numerous speakers. Additionally, we propose SS-MoE, a hybrid mixture-of-experts model that combines shared experts and soft routing. Integrated into NSD-MS2S, the resulting NSD-MS2S-SSMoE system surpasses the baseline and shows competitive performance on the CHiME-7 evaluation set. Together, these contributions provide an effective and robust solution for speaker diarization in complex real-world environments.

\bibliographystyle{IEEEtran}
\bibliography{mybib}

\vfill

\end{document}